\documentclass[a4paper,12pt]{article}
\usepackage[T1]{fontenc}
\usepackage[english]{babel}
\begin{document}

\title{ Effective Lagrangian for  $SU(3)_c$ gluodynamics
from  lattice color dielectrics }

\author{by \\
H. Arod\'z   \\
Institute of Physics, Jagellonian University\\
Reymonta 4, 30-059 Cracow, Poland \\
and \\
H.-J. Pirner \\ Institute of Theoretical Physics, 
Ruprecht-Karls University\\  Philosophenweg 19,
Heidelberg, Germany} 
\date{ }
\maketitle
\thispagestyle{empty}

\begin{abstract}
Small momenta limit of the lattice color dielectric model in the $SU(3)_c$ 
case is considered. We show that in that limit the model involves a 
coarse-grained gluon field, a  color singlet "bleached gluon" field, and 
both color singlet and octet rank two symmetric tensor fields constructed 
from the first order derivatives of a vector field. In contradistinction 
from other approaches we do not obtain scalar color dielectric fields 
directly, but they can be introduced as traces of the tensor fields.
\end{abstract}

\pagebreak

\setcounter{page}{1}

\section{Introduction}
At present there exist several theoretical approaches to hadron
physics at low energies. The lattice QCD provides a microscopic, {\it ab
initio} approach. Another class of approaches is based on phenomenological 
models in continuum space-time. Such models describe main features of the 
hadron spectrum within relatively simple formalism ---  the bag model or 
the skyrmion model are good examples. 

Friedberg and Lee \cite{1} proposed a phenomenological model 
with a dynamical non-Abelian gauge field (coarse-grained gluon field),
which is interpreted as a small momentum
component of the original gluon field of QCD. The model also involves 
a color singlet  scalar field coupled to the gauge field. 
Its role is to mimic effects of quantum
fluctuations of the gluon field with momenta larger than or of the order of
the confinement scale.  The model has been refined and succesfully
applied in calculations of static hadron characteristics. 
For review, references and examples of recent applications see, 
e.g., \cite{2}, \cite{3}, \cite{4}, \cite{5}.  The Friedberg-Lee model is 
particularly interesting because of the presence of the non-Abelian gauge 
field -- it takes into account the low momentum glue component of hadrons.

Long time ago  Nielsen and Patkos \cite{6} have pointed out that the 
Friedberg-Lee model \footnote{ We call it also "the continuum color 
dielectric model".} appears naturally within QCD framework if we consider
the gluon field averaged over a small space-time cell.
In 1984 Mack \cite{7} introduced a lattice version of the color dielectric 
model. Later on, in a series of papers, see \cite{8}, \cite{9}, \cite{10} and 
references therein, Pirner and collaborators attempted to derive the lattice 
color dielectric model from the "microscopic" lattice QCD -- the results
were quite encouraging. The lattice color dielectric model was studied with 
the help of usual lattice techniques, essentially in the strong
coupling regime. Recent presentation of the idea of the lattice color
dielectric model and  related results on 2+1 dimensional Yang-Mills theory
can be found in \cite{11}. As pointed out in that paper, the color 
dielectric model can be labelled as a "simple vacuum - complicated action"
approach, while other approaches to QCD are rather of the type 
"simple action - complicated vacuum". 
Here "simple action" or "complicated action" refers mainly to
the field content --- in the "simple" case only the standard $SU(3)_c$
gauge field is present, while the color dielectric action involves a wider 
class of fields. 

The investigations of the lattice color dielectric model have shown 
that apart from the coarse-grained gluon field several other fields are
involved, not merely the single color singlet scalar field.
In $SU(2)_c$ case four real, color singlet fields appear. In the
$SU(3)_c$ case there is a  color singlet vector field (called in \cite{10} 
and here the bleached gluon), and four Hermitean, $3\times3$, positive 
definite matrix fields. These lattice results suggest that the Friedberg-Lee 
type models can be modified accordingly. In the present work we take first 
steps in this direction.

We consider the small momenta limit of the gluonic part of the lattice color 
dielectric model.  In this limit the continuum space-time description is 
the most appropriate one, and we would like to find the resulting field 
theoretical model. Obtained in this way gluonic effective Lagrangian for 
fields on the continuum space-time will be distinguished by its
close connection with the lattice QCD.

The small momenta limit of the lattice $SU(3)_c$ color dielectric model
was already considered in \cite{10} with the help of strong simplifying 
assumptions, which reduced the Hermitean matrix fields to one
color singlet scalar field. The main problems with keeping the full 
set of fields were:
(i) how to find the continuum fields corresponding to the Hermitean matrix
fields, (ii) how to ensure $SO(4)$ invariance of the 
Euclidean continuum effective action.
In the present paper we propose solutions to these two problems.
With the help of certain rather nontrivial field transformation we find 
continuum counterparts for the Hermitean matrix fields. It turns out that 
in the $SU(3)_c$ case the coarse-grained gluon field
is accompanied by a rank two symmetric tensor field constructed from a 
vector field.  This is one of the main results of our work. 
The bleached gluon field found already in
\cite{10} also is present. The tensor field can be decomposed into $SO(4)$
scalar and traceless symmetric parts. The color singlet part of the
scalar component can be identified with the scalar color dielectric field 
present in the Friedberg-Lee model.
Thus, the small momenta limit of the lattice $SU(3)_c$ color dielectric 
model yields rather large set of color dielectric fields: the $SU(3)_c$ 
singlet and octet components of the $SO(4)$ scalar and (rank two) tensor. 

In order to solve the problem with $SO(4)$ invariance we use an averaging 
over all $SO(4)$ orientations of the lattice. This simple procedure
ensures the $SO(4)$ invariance of the continuum effective action.

The plan of our paper is as follows. In Section 2, which has introductory
character,  we describe basic ingredients of the lattice color dielectric 
model. In Section 3 we propose the field transformation, and we introduce 
the continuum fields corresponding to the lattice color dielectric fields. 
Effective action of the lattice color dielectric model and its small momenta
limit are discussed in Section 4. In Section 5 we have collected remarks 
related to our work. The Appendix contains mathematical results pertinent 
to the averaging over the $SO(4)$ group.

\section{The lattice color dielectric fields}
The lattice color dielectric model can be described as a lattice gauge
theory in which the standard unimodular link variables
$U_{k}(x)\in SU(3)_c$ are replaced by  more general 3 by 3 matrices \cite{7}
\begin{equation}
U_{k}(x) \rightarrow \Phi_{k}(x).
\end{equation}
Here $x$ denotes sites, and $k=1,2,3,4$ enumerates positive direction links 
of a four dimensional hypercubic lattice in the Euclidean space-time M. 
The four links start at the point $x$.
In papers \cite{7,8} the effective non-unitary link variables
$\Phi_k(x)$ are introduced as weighted sums of products of 
"microscopic" unimodular link variables, 
defined on another lattice, finer (i.e., with shorter links) 
than the one introduced above. A ``blocking" procedure leads from the 
original lattice QCD defined on the fine lattice to the color 
dielectric model on the coarser lattice.  The hope is that the "macroscopic"
link variable $\Phi_k(x)$ is infrared stable, that is its behaviour at
small momenta is not dominated by quantum fluctuations. 
The corresponding weighting function is normalised so that a 
matrix norm of $\Phi_k(x)$ is not greater than 1. The coarse lattice link 
length is denoted by $b$, and it is assumed to be large enough for the 
infrared stability. For concreteness, we expect that $b\sim 0.4$ fm, but in
the present paper we do not attempt to give a well-founded estimate. 

Upon inversion of the direction of the link  \cite{7}
\begin{equation}
\Phi_{-k}(x+b e_k) = \Phi^{\dagger}_k(x).
\end{equation}
In this formula $e_{k}$ is
the unit four-vector corresponding to the positive direction link $k$, 
and $-k$ denotes the link obtained from the link $k$ by reversing its 
direction. The link $-k$ in formula (2) starts at the point $x+be_k$ and 
it ends at the point $x$. Under the $SU(3)_c$ gauge transformations the 
effective link variables $\Phi_k(x)$ transform in the usual manner, 
\begin{equation}
\Phi'_k(x) = \omega(x+b e_k) \Phi_k(x) \omega(x)^{-1},
\end{equation}
where $\omega \in SU(3)_c$. 
From the point of view of the continuum QCD, these gauge 
transformations are the coarse-grained ones,
that is the gauge function $\omega$ is approximately constant over 
distances smaller than a fraction of $b$.

Action functional $S[\Phi]$ of the lattice color dielectric model is
built of gauge invariant products of $\Phi$'s.
The main idea underlying the
color dielectric model is that if the coarse lattice link length $b$ 
is large enough,  then the vacuum value of $\Phi_k$ should vanish. This
requirement restricts the form of the action.
We will discuss the action functional in Section 4. Before that, 
in Section 3, we will analyse field contents of the model in the small momenta
limit.

Following the previous approaches, we base our analysis of the 
field content of the lattice color dielectric model  on the
polar decomposition of the matrices $\Phi_k(x)$:
\begin{equation}
\Phi_k(x) = exp(i \theta_{k}(x)) V_{k}(x) \hat{\chi}_{k}(x),
\end{equation}
where $V_{k}(x)$ is a matrix of the $SU(3)$ type, 
$\hat{\chi}_{k}(x) = \left(\Phi^{\dagger}_k(x) \Phi_k(x)\right)^{1/2}$ is a  
3 by 3 Hermitean matrix with nonnegative eigenvalues, 
and  $\theta_{k}(x)$ is a real number.

The polar decomposition (4) is analogous to a transformation from Cartesian
to spherical coordinates, with $\theta_k$ and $\hat{a}_k$ (introduced in 
formula (5) below) playing the role of angle variables, 
while $\hat{\chi}_k$ corresponds to the radius. Such transformation 
has the drawback: it is singular at $\hat{\chi}_k =0$. It is clear from
formula (4) that $\theta_k(x)$ and $V_k(x)$ can be taken arbitrary if
$\hat{\chi}_k =0$. This means that the vacuum values of the field 
$\theta_{\mu}$ and of the coarse-grained gluon field $\hat{A}_{\mu}$  can
be arbitrary. \footnote{ In a sense this agrees with the picture 
arising in the mentioned in the Introduction "simple action - 
complicated vacuum" approach, where 
low momenta components of the non-Abelian  gauge field also do not have 
preferred vacuum values because they strongly fluctuate.}
On the other hand, with the polar decomposition (4) we have a smooth
correspondence with the original $SU(3)_c$ gauge field theory, which is
obtained when $\hat{\chi}_k = I$ and $ \exp(i\theta_k(x)) = 1$ --- in this
region the polar decomposition is nonsingular.

$V_{k}$ can be related in the standard manner 
to a traceless Hermitean gauge field $\hat{A}_{\mu}$ on M:
\begin{equation}
V_{k}(x) = exp[i\hat{a}_{k}(x)],
\end{equation}
where $\hat{a}_{k}(x)$ is regarded as the projection of the continuum 
gauge field $\hat{A}_{\mu}(x) $ on the link four-vector $b e_{k}$  
(pointing in the direction $k$ at the site $x$), that is
\begin{equation}
\hat{a}_{k}(x) = b e^{\mu}_k \hat{A}_{\mu}(x).
\end{equation}
The Greek indices refer to the continuum Euclidean 
space-time $M$ in which the original (``fine grid") as well as
the coarse lattices are constructed. The $\hat{A}_{\mu}(x)$ field is 
called the coarse-grained (or averaged) $SU(3)_c$ gauge field.  Similarly,
$\theta_{k}(x)$ is related to a color singlet vector field $\Theta_{\mu}$ 
defined on $M$
\begin{equation}
\theta_{k}(x) = b e^{\mu}_k \Theta_{\mu}(x).
\end{equation}
The $\Theta_{\mu}(x)$ field is called in \cite{10} and in the following
the bleached gluon. 

The formulas presented above  imply that  $V_k, \theta_k$ are scalars with
respect to the $SO(4)$ rotations in $M$. The non-unitary link variables 
$\Phi_k(x)$ are $SO(4)$ scalars too, because they are defined as the 
weighted sums of products of the microscopic link variables of the original 
QCD. The microscopic link variables -- being related to projections of the
original microscopic gauge field onto the link four-vectors of the
initial fine lattice (the corresponding formulas are analogous
to formulas (5), (6) above) -- are  $SO(4)$ scalars.   
Because  $\Phi_k, V_k, \theta_k$ in formula (4) are  $SO(4)$ scalars,
it follows that also $\hat{\chi}_{k}$ are $SO(4)$ scalars.

By assumption,  $V_k$ transforms under the $SU(3)_c$ gauge transformations 
as thelattice $SU(3)_c$ gauge field -- a formula analogous to (3) -- and 
$\theta_k$ is gauge invariant. Then, it follows from (4) that $\hat{\chi}_k$
belongs to the adjoint representation of $SU(3)_c$
\begin{equation}
\hat{\chi}'_k(x) = \omega(x) \hat{\chi}_k(x) \omega(x)^{-1}.
\end{equation}

Formulas (5), (6), (7) imply that
\begin{eqnarray}
& V_{-k}(x+b e_k) = V_{k}^{\dagger}(x), & \\
& exp[i\theta_{-k}(x+b e_k)] = exp[-i\theta_{k}(x)],
\end{eqnarray}
and formulas (2), (4) give
\begin{equation}
\hat{\chi}_{-k}(x+b e_k) = V_{k}(x)\hat{\chi}_{k}(x)V_{k}^{\dagger}(x).
\end{equation}

In the $SU(2)_c$ case the $\theta_k$ field is absent, and each 
matrix $\hat{\chi}_k(x)$ is replaced by a nonnegative number $\chi_k(x)$, 
see \cite{7,9}, invariant under the $SU(2)_c$ gauge transformations.

\section{ The field transformation }
We are interested in the small momenta sector of the
lattice color dielectric model, specifically, we would like to obtain the
continuum space-time formulation of the model in that sector. 
The main problem here is to find the continuum counterpart
for the lattice  fields $\hat{\chi}_k(x)$. 
The properties of these fields are rather puzzling. By construction, 
$\hat{\chi}_k(x)$ are defined on the links, like the lattice gauge 
field. In spite of that they cannot be related to a gauge field
on $M$ because then the gauge transformation law would have to have 
the form (3), and not (8).
The transformation (8) is typical for a matter field defined on the sites 
of the lattice, but $\hat{\chi}_k(x)$ is defined on the links.  
To solve this problem we use a field transformation which relates the 
$\hat{\chi}_k(x)$ fields to components of a Hermitean vector field 
$\hat{B}_{k}(x),$  $k=1,2,3,4$, located on the sites of the lattice. Namely,
\begin{equation}
\hat{\chi}_k(x) = G \left( D_k  \hat{B}_k(x) \right)
\end{equation}
(no summation over $k$). Here  $G(\cdot)$ denotes a matrix function discussed 
below, and 
\begin{equation}
D_l\hat{B}_k(x) = V^{\dagger}_l(x) \hat{B}_k(x+b e_l) V_l(x) - \hat{B}_k(x).
\end{equation}
is the lattice version of gauge-covariant derivative. The 
$\hat{B}_{k}(x)$ field has the following  $SU(3)_c$ gauge transformation
\begin{equation}
\hat{B}'_{k}(x) = \omega(x) \hat{B}_{k}(x) \omega(x)^{-1}.
\end{equation}
Thus, $\hat{B}_k(x)$ is a matter field. 

Formula (12) is consistent with the properties (8) and (11):
the transformation law (8) for
$\hat{\chi}_k(x)$ fields follows immediately from formulas (12-14), and
in order to obtain relation (11) it is enough to notice that
\begin{equation}
\hat{B}_{-k}(x) = - \hat{B}_k(x),
\end{equation}
and
\[ 
D_{-k}\hat{B}_{-k}(x +b e_k) = V_k(x) D_k\hat{B}_k(x) V^{\dagger}_k(x).
\]
Formula (15) follows from the assumption that $\hat{B}_k(x)$ is the projection
of a continuum vector field $\hat{B}_{\nu}(x)$ on the link four-vector,
see formula (20) below. 

The function $G(\cdot)$ is supposed to transform Hermitean matrices into  
Hermitean ones, hence coefficients of 
its Taylor expansion should be real numbers. Moreover, its matrix values
should be positive definite because $\hat{\chi}_k(x)$ in formula 
(12) is positive definite. Finally,  classical vacuum value of the 
$\hat{\chi}_k(x)$ fields is expected to vanish in the color dielectric 
models.  If we require that the corresponding vacuum value of 
$D_k\hat{B}_k(x)$ also vanishes, then we may take
\begin{equation}
G(DB) = \left( c_1 DB + c_2 (DB)^2 + ... \right)^2,
\end{equation}
where $DB$ is a shorthand for $ D_k\hat{B}_k(x)$, and all $c_i$ are real. 
For $DB$ close to its vacuum value, that is when 
\begin{equation}
D_k\hat{B}_k(x) \ll 1,
\end{equation}
we may neglect the higher powers of $DB$, and then
\begin{equation}
\hat{\chi}_k(x) = \left( D_k\hat{B}_k(x)\right)^2.
\end{equation}
The constant $c_1$ has been removed by rescaling $\hat{B}_k(x)$.
Below we shall use the field transformation (18).

Now, let us consider the small momenta limit. The lattice field $V_k(x)$, 
formulas (5), (6), is now associated with the exponential
\begin{equation}
V_k(x) = P exp( ib \int \hat{A}_{\mu}(y) dy^{\mu}),
\end{equation}
path ordered along the straightline segment $[x, x+b e_k]$ in the continuum
space-time $M$. Here $\hat{A}_{\mu}(y)$ is the coarse-grained gauge field 
on the continuum Euclidean space-time $M$ -- it is a continuous function of 
$y$, almost constant on the distance $b$. Formula (19) gives (5), (6) in the 
leading order in $b$. Our definitions imply that the continuum $SU(3)_c$ gauge 
transformation of the coarse-grained gauge field has the form
\[ 
\hat{A}'_{\mu} = \omega \hat{A}_{\mu} \omega^{-1} - i \partial_{\mu}\omega
\omega^{-1}.
\]

The lattice fields $\hat{B}_{k}(x)$  in the small momenta sector
are regarded as projections (on the lattice four-vectors $be_k$)
of the vector field $\hat{B}_{\nu}(x), $ which is defined on $M$ and 
almost constant on distances $b$, 
\begin{equation}
\hat{B}_k(x) = b e_k^{\nu} \hat{B}_{\nu}(x).
\end{equation}
The lattice covariant derivative $D_k$ is interpreted as
\[ D_k = b e_k^{\mu} D_{\mu},
\]
where $D_{\mu}$ is the covariant derivative in the continuum
space-time $M$. \footnote{ Due to the presence of $b$ in the last two 
formulas, $D_{\mu}$ and $\hat{B}_{\nu}$ have the usual dimension $cm^{-1}$. }
Then, formulas (13), (19) give
\begin{equation}
D_k \hat{B}_{k}(x) = b^2 e^{\mu}_k e^{\nu}_k D_{\mu}\hat{B}_{\nu}(x)
+  b^3 e^{\mu}_k e^{\sigma}_k e^{\nu}_k 
D_{\mu} D_{\sigma} \hat{B}_{\nu}(x) + ... ,
\end{equation}
where 
\[
D_{\mu}\hat{B}_{\nu} = \partial_{\mu}\hat{B}_{\nu} - i [ \hat{A}_{\mu} ,
\hat{B}_{\nu}].
\]
Let us note that $\hat{B}_{\nu}$ has both color singlet and octet components; 
for the singlet component $D_{\mu}\hat{B}_{\nu}$ reduces to the ordinary 
derivative. 

Thus, in the small momenta limit 
\[
D_k\hat{B}_k \cong b^2 e_k^{\mu} e_k^{\nu} D_{\mu}\hat{B}_{\nu}
\]
(no summation over $k$). This formula implies that $D_k\hat{B}_k$ depends 
only on the symmetric part of $D_{\mu}\hat{B}_{\nu}$,
\begin{equation}
D_k\hat{B}_k(x) \cong b^2 e_k^{\mu} e_k^{\nu} \hat{G}_{\mu\nu}(x),
\end{equation}
where 
\begin{equation}
\hat{G}_{\mu\nu}(x) = \frac{1}{2} \left( D_{\mu}\hat{B}_{\nu}(x) + 
D_{\nu}\hat{B}_{\mu}(x)\right) = \hat{G}_{\nu\mu}(x).
\end{equation}
It is clear that under the $SU(3)_c$ gauge transformations
\[
\hat{G}'_{\mu\nu}(x) = \omega(x) \hat{G}_{\mu\nu}(x) \omega(x)^{-1}.
\]
Finally, considering formulas (18) and (22) we conclude that in the small 
momenta limit 
\begin{equation}
\hat{\chi}_k(x) \cong b^4 e_k^{\mu}e_k^{\nu}e_k^{\rho}e_k^{\lambda}
\hat{G}_{\mu\nu}(x) \hat{G}_{\rho\lambda}(x)
\end{equation}
(no summation over $k$).

The field $\hat{B}_{\mu}$ enters only through
the symmetrized covariant derivatives, that is through $\hat{G}_{\mu\nu}$.
Therefore, in all calculations in the present paper we may use just the
$\hat{G}_{\mu\nu}$ field. Nevertheless, the basic
dynamical field in the small momenta limit is the $\hat{B}_{\mu}$ field. This
makes difference when, for example, deriving field equations because
variational derivative of the action  should be taken
with respect to $\hat{B}_{\mu}$ and not $\hat{G}_{\mu\nu}$.

The tensor field $\hat{G}_{\mu\nu}$  can be decomposed into two 
parts which are irreducible with respect to the $SO(4)$ group, namely
\begin{equation}
\hat{G}_{\mu\nu}(x) = \hat{\sigma}(x) \delta_{\mu\nu} +\hat{g}_{\mu\nu}(x),
\end{equation}
where $(\hat{g}_{\mu\nu})$ has vanishing trace, $\hat{g}_{\mu\mu}=0$.
Formulas (23) and (25) imply that 
\[
\hat{\sigma}(x) = \frac{1}{4} D_{\mu} \hat{B}_{\mu}(x).
\]
Because $e^{\mu}_k e^{\mu}_k =1$ for each $k=1,2,3,4,$ formulas (24) and 
(25) give 
\[
\hat{\chi}_k(x) = b^4 \left[ \hat{\sigma}^2(x) + e^{\mu}_k e^{\nu}_k \left(
\hat{\sigma}(x)\hat{g}_{\mu\nu}(x) + \hat{g}_{\mu\nu}(x) \hat{\sigma}(x)
\right) + e^{\mu}_k e^{\nu}_ke^{\rho}_k e^{\lambda}_k \hat{g}_{\mu\nu}(x)
\hat{g}_{\rho \lambda}(x) \right]
\]
(no summation over $k$). 
Furthermore, $ \hat{\sigma}$ and $\hat{g}_{\mu\nu}$ can be split into the 
color singlet and octet parts,
\[
\hat{\sigma}(x) = \sigma(x) I + \frac{1}{2} \lambda_a \sigma^a(x), \;\;
\hat{g}_{\mu\nu}(x) =
g_{\mu\nu}(x) I + \frac{1}{2} \lambda_a g^a_{\mu\nu}(x),
\]
where $\lambda_a$ are the Gell-Mann matrices. The fields 
$\sigma$, $\sigma^a$, $g_{\mu\nu}$ and $g_{\mu\nu}^a$ are real due to 
hermicity of $\hat{G}_{\mu\nu}$.

In the $SU(2)_c$ case only the color singlet parts are present, that is 
$G_{\mu\nu} = \sigma(x)  \delta_{\mu\nu} + g_{\mu\nu}(x)$. 
 
\section{Towards the continuum color dielectric action}
We have seen that in the small momenta limit the lattice color dielectric 
model can be formulated in terms of the fields $\hat{A}_{\mu}, 
\hat{B}_{\mu}, \Theta_{\mu},$ which are defined on the 
continuum space-time. Now we would like to introduce certain  lattice color 
dielectric action, and to discuss its small momenta limit. Our ultimate goal
is to find the corresponding continuum effective Lagrangian for the fields  
$\hat{A}_{\mu}, \hat{B}_{\mu}, \Theta_{\mu}.$  The  discussion  presented 
below has rather preliminary character. We concentrate on theoretical aspects,
in particular on the problem of restoration of the $SO(4)$ symmetry broken in 
the lattice model.

Let us start from the following remark. The lattice color dielectric 
model has been considered in the papers \cite{7,8,9,10,11} as an example
of quantum lattice gauge theory with the matrix quantum variables 
$\Phi_k(x)$ and with certain action $S[\Phi]$. The functional integration 
measure is given by  $[d\Phi] = \prod_{x,k}d\Phi_k(x)$. This quantum theory 
can be described in terms of the corresponding effective action $S_c$ 
calculated in a standard way, e.g., with the help of background field method.
In the tree (mean field) approximation the effective action 
coincides with the latttice color dielectric action $S[\Phi]$, but its full 
form by definition summarizes all quantum corrections. The lattice color 
dielectric action $S_c$ which we consider 
below is meant as an Ansatz for such effective action. Therefore, it is not 
necessary to discuss here the small momenta limit of the functional integral, 
e.g., of the measure $[d\Phi]$. In principle, the precise form of the lattice 
color dielectric effective action $S_c$ follows from the underlying 
microscopic lattice QCD. Nevertheless, it seems more practical to assume its 
form and to fit unknown parameters like in other effective models, see 
\cite{11} for a lucid discussion of this point. The action $S_c$ 
is assumed to be invariant with respect to the coarse-grained $SU(3)_c$ gauge 
transformations. Moreover, the  terms with higher powers of $\Phi_k$ are 
expected to be less important because the matrix norm of $\Phi_k$ is smaller
than 1. 

Below we present an Ansatz for the lattice color dielectric effective action
$S_c$. It satisfies the following general requirements: the $SU(3)_c$ gauge 
invariance, positivity, absolute minimum for $\Phi_k =0$. Moreover, it 
involves low powers of $\Phi_k$. 

The first contribution to the action $S_c$ has the following form
\begin{equation}
S_1[\Phi] = \sum_{x,k,l} Tr \left( \Psi_{k,l}^{\dagger}(x) 
\Psi_{k,l}(x)\right), 
\end{equation}
where  $k=1,...,4$ and $l=\pm1,...,\pm4$.
(If a link with negative $l$ starts at the point $x$ then it ends at the 
point $x - b e_{|l|}$.)  $\;\Psi_{k,l}(x)$ is a gauge covariant polynomial in 
$\Phi_k(x)$, that is under the $SU(3)_c$ gauge transformations
\[
\Psi'_{k,l}(x) = \omega(x + be_k) \Psi_{k,l}(x) \omega^{-1}(x).
\]
Specifically, we shall take
\begin{equation}
\Psi_{k,l}(x) = \Phi_k(x) +  \alpha_1
\Phi^{\dagger}_l(x+be_k) \Phi_k(x+be_l) \Phi_l(x),
\end{equation}
where $\alpha_1$ is a real constant. The second term on the r.h.s. of 
formula (27) corresponds to a path of length $3b$ connecting the points $x$ 
and $x+be_k$. For $l\neq \pm k$ the path has the staple-like shape. 
It is clear that if the constant $\alpha_1$ is not too large, all 
$\Psi_{k,l}$ vanish only when all $\Phi_k$ vanish -- this ensures that 
$S_1[\Phi]$ has the absolute, nondegenerate minimum for $\Phi_k =0$.

Let us introduce the following notation
\begin{equation}
\hat{h}_{k,l}(x) = V^{\dagger}_l(x) \hat{\chi}_k(x+be_l) V_l(x).
\end{equation}
Adding $\hat{\chi}_k(x) -\hat{\chi}_k(x)$ to the r.h.s. of formula (28) we 
see that
\begin{equation}
\hat{h}_{k,l}(x) = \hat{\chi}_k(x) + D_l\hat{\chi}_k(x).
\end{equation}
We also introduce the field strength $\hat{F}_{kl}$ of the lattice 
coarse-grained $SU(3)_c$ gauge field $V_k$
\begin{equation}
\exp(i \hat{F}_{kl}) =
V_k^{\dagger}(x) V^{\dagger}_l(x+be_k) V_k(x+be_l) V_l(x).
\end{equation}
Similarly, we define  $f_{kl}$ -- the Abelian field strength corresponding 
to the lattice $\theta_k$ field,
\begin{equation}
\exp(i f_{kl}) =
\exp[i (  \theta_k(x+be_l) + \theta_l(x) - \theta_k(x) -
\theta_l(x+be_k) ) ]. 
\end{equation}
With this notation we may write
\begin{eqnarray}
&  \Psi_{k,l}(x) = \exp(i\theta_k(x)) V_k(x) \left[\hat{\chi}_k(x) \right. & 
\nonumber  \\
& \left.  +  \alpha_1 \hat{h}_{l,k}(x) \exp(i\hat{F}_{kl}(x) 
+ i f_{kl}(x)I) \hat{h}_{k,l}(x) \hat{\chi}_l(x)\right], &
\end{eqnarray}
where $I$ denotes the 3 by 3 unit matrix. The factor 
$\exp(i\theta_k(x))V_k(x)$ cancels out in the product 
$\Psi^{\dagger}_{k,l}\Psi_{k,l}$ in formula (26). 

In the case of weak fields $\hat{F}_{kl}, f_{kl}, D_l\hat{\chi}_k$,
the first nontrivial approximation to the expression in the square bracket 
on the r.h.s. of formula (32) has the form 
\begin{eqnarray}
& [ ... ] \cong  \hat{\chi}_k(x) + \alpha_1 \left[
\hat{\chi}_l(x)\hat{\chi}_k(x) \hat{\chi}_l(x)  +
D_k\hat{\chi}_l(x)\hat{\chi}_k(x) \hat{\chi}_l(x) \right.  & \nonumber \\
& \left. +  \hat{\chi}_l(x)D_l\hat{\chi}_k(x)\hat{\chi}_l(x) 
+ i \hat{\chi}_l(x) \left(\hat{F}_{kl} + f_{kl} I \right)
\hat{\chi}_k(x) \hat{\chi}_l(x) \right]. &
\end{eqnarray}

The lattice color dielectric effective action can not be just equal to 
$S_1[\Phi]$ because in that case it would have too large symmetry group --
$S_1[\Phi]$ is invariant under $U(3)_c$  gauge transformations. In order 
to break this symmetry down to the $SU(3)_c$  gauge symmetry 
we use $Det \Phi_k$ which is invariant under the $SU(3)_c$ gauge 
transformations only. The polar decomposition (4) gives
\[ Det \Phi_k = \exp(3i\theta_k(x)) Det  \hat{\chi}_{k}. \] Let us notice 
that $ Det \hat{\chi}_{k}$ can be expressed by traces of powers of
$ \hat{\chi}_{k}$. For this one may use  Hamilton-Cayley identity, 
which for arbitrary 3 by 3 matrix $\hat{C}$ has the following form
\[ \hat{C}^3 - Tr\hat{C}\; \hat{C}^2 +\frac{1}{2} [(Tr\hat{C})^2 - 
Tr(\hat{C}^2)] \hat{C} - Det\hat{C}\; I =0. \]
Taking trace of this identity and putting $\hat{C}=  \hat{\chi}_{k}$ we find 
that 
\[ Det  \hat{\chi}_{k} = \frac{1}{3} Tr( \hat{\chi}_{k}^3) - \frac{1}{2} 
Tr \hat{\chi}_{k} Tr( \hat{\chi}_{k}^2) + \frac{1}{6} (Tr\hat{\chi}_{k})^3.\]
Because $Det \Phi_k$ can be a complex number it cannot be directly included 
into the action. Instead, we may take
\begin{equation}
S_2[\Phi] = \gamma \sum_{x,k} \left(Im Det \Phi_k(x)\right)^2 =
\gamma \sum_{x,k} \left( Det\hat{\chi}_k(x) \right)^2 \sin^2(3\theta_k(x)),
\end{equation}
where $\gamma$ is a positive constant. 
For $\theta_k \ll 1$ we may approximate $\sin^2(3\theta_k) \cong 9 b^2
e^{\mu}_ke_k^{\nu} \Theta_{\mu}(x)\Theta_{\nu}(x).$

We may also include in $S_c$ terms which contain powers of 
$\hat{\chi}_k$ only. For example, up to the fourth power in $\hat{\chi}_k$ 
\begin{eqnarray}
& S_p = \lambda_2 \sum_{x,k} Tr\left(\Phi^{\dagger}_k(x) \Phi_k(x)\right) 
& \nonumber \\ 
& + \lambda_3 \sum_{x,k,l} Tr\left(\Phi^{\dagger}_k(x) 
\Phi_k(x)\Phi^{\dagger}_l(x)
\Phi_l(x)\right) &  \\
& +  \lambda_4\sum_{x,k,l} Tr\left(\Phi^{\dagger}_k(x) 
\Phi_k(x)\right) Tr\left(\Phi^{\dagger}_l(x)\Phi_l(x)\right) & \nonumber 
\end{eqnarray}
where $\lambda_i$ are positive constants. It is easy to see that 
$\exp(i\theta_k)$ and $V_k$ cancel out in each term on the r.h.s. of 
this formula -- only $\hat{\chi}_k$'s are left. 

The total lattice color dielectric effective action is given by the sum of 
these contributions,
\[ S_c= S_1 +S_2 +S_p.  \]

In the small momenta limit $\hat{F}_{kl}$ and $f_{kl}$ are related to the 
projections on the lattice four-vectors of the corresponding 
tensors in the continuum space-time,
\begin{equation}
\hat{F}_{kl} \cong b^2 e^{\mu}_k e_l^{\nu} \hat{F}_{\mu\nu}, \;\;\;
f_{kl} \cong b^2 e^{\mu}_k e_l^{\nu} f_{\mu\nu},
\end{equation}
where $\hat{F}_{\mu\nu} = \partial_{\mu}\hat{A}_{\nu} - \partial_{\nu}
\hat{A}_{\mu} - i [\hat{A}_{\mu} , \hat{A}_{\nu}] $ and
$ f_{\mu\nu} = \partial_{\mu} \Theta_{\nu} - \partial_{\nu} \Theta_{\mu}. $
For $\hat{\chi}_k$ we have formula (24), and
\begin{equation}
D_l\hat{\chi}_k \cong b^5
e^{\mu}_l e^{\nu}_ke^{\lambda}_ke^{\sigma}_ke^{\rho}_k
\left( D_{\mu}\hat{G}_{\nu\lambda}(x) \hat{G}_{\sigma\rho}(x)
+ \hat{G}_{\nu\lambda}(x) D_{\mu}\hat{G}_{\sigma\rho}(x) \right),
\end{equation}
as follows from (24) by Leibniz rule for the covariant derivative of a 
product.

Inserting formulas (36), (37) on the r.h.s. of formulas (26), (32), (34), 
(35) for $S_1, S_2, S_p,$ and replacing $\sum_x$ by $b^{-4} \int d^4x$, we 
obtain the action functional
$S_c[\hat{A}_{\mu}, \Theta_{\nu}, \hat{G}_{\mu\nu}]$ of the effective  model 
in the Euclidean continuum space-time. It is clear that all terms in $S_c$  
have the form of contractions of multiple factors $\Theta_{\mu}, 
\hat{G}_{\mu\nu}, D_{\mu}\hat{G}_{\nu\lambda}, \hat{F}_{\lambda \sigma}, 
f_{\rho\omega}$ with the products 
$e^{\mu}_ke^{\nu}_ke^{\rho}_ke^{\lambda}_ke^{\alpha}_l
e^{\beta}_l...     $
of components of the lattice unit four-vectors  $e_k$.
Such products are lattice artifacts, and they explicitly break the 
Euclidean $SO(4)$ invariance. The contractions give $SO(4)$ scalars, 
but their form in general is not invariant with respect to the $SO(4)$ 
transformations. Thus, the action  
$ S_c[\hat{A}_{\mu}, \Theta_{\nu}, \hat{G}_{\mu\nu}]$ is not 
$SO(4)$ invariant, eventhough it is  $SO(4)$ scalar. This is the second 
problem mentioned in the Introduction.

To solve this problem, we proceed as follows.
The four-vectors  $e_k, k=1,...,4,$ form an orthonormal basis in the
Euclidean space-time M. This basis does not have any physical meaning and it 
can be arbitrary --- the lattice can have any orientation with respect to a
"laboratory" reference frame in M. Nevertheless, each concrete choice and the
subsequent lattice formulation break the $SO(4)$ invariance.  We can restore 
this symmetry by integrating  over all $SO(4)$ orientations of 
the basis. Formally, this can be achieved by preceding each term in the
effective action by the normalized to unity Haar integral 
$\int_{SO(4)} d {\cal O}$  over the $SO(4)$ group, and by regarding the 
four four-vectors  $e_k$ as (orthonormal) columns of the matrix 
${\cal O}\in SO(4)$, that is
\[  
e_k^{\mu} = {\cal O}_{k\mu}.
\]

Due to the $SO(4)$ invariance of the Haar measure, the integrals are
forminvariant with respect to simultaneous $SO(4)$ transformations of all
indices $\mu, \nu, \rho$, etc., which refer to the "laboratory" 
reference frame in the continuous Euclidean space-time $M$.
For example, the term $Tr\left(\Phi^{\dagger}_k(x) \Phi_k(x)\right)$ 
acquires the form
\[
\eta_{\mu_1 \mu_2 ... \mu_8} b^{8}  \;
Tr \left( \hat{G}_{\mu_1\mu_2}(x) \hat{G}_{\mu_3\mu_4}(x)
\hat{G}_{\mu_5\mu_6}(x) \hat{G}_{\mu_7\mu_8}(x) \right),
\]
where
\[
\eta_{\mu_1 \mu_2 ... \mu_8} =
\int_{SO(4)} d{\cal O} {\cal O}_{k\mu_1} ... {\cal O}_{k\mu_8}
\]
(no summation over k). It is easy to see that $\eta_{\mu_1 \mu_2 ... \mu_8}$  
does not depend on $k$, and that it is
forminvariant under simultaneous $SO(4)$ transformations of all 
its indices. Because $\eta_{\mu_1 \mu_2 ... \mu_8}$ is symmetric with 
respect to permutations of its indices, it is equal to a sum of 
products of Kronecker deltas $\delta_{\mu_i \mu_j}$, where $i,j = 1...8.$
In general case we encounter integrals of the type
\[
\int_{SO(4)} d{\cal O} e^{\mu_1}_k...e_k^{\mu_m}
e_l^{\nu_1}...e_l^{\nu_n},
\]
where $ e_k^{\mu} = {\cal O}_{k\mu} , \;\; e_l^{\nu} = {\cal O}_{l\nu}$.
They can be calculated with the help of method presented 
in the Appendix. The  result has the form of a sum  of products of 
Kronecker deltas with  permuted indices  $\mu_i, \nu_j$.

Because the $SO(4)$ integral of any product of odd number of the matrix
elements ${\cal O}_{k\nu}$ vanishes (see the Appendix), some terms 
in  $S_1[\hat{A}, \Theta, \hat{G}]$ vanish.
Other terms vanish due to the sum over $l = \pm1,...,\pm4$,
because $\hat{F}_{-lk}(x) = - \hat{F}_{lk}(x), \; D_{-l}\hat{\chi}_k(x) = -
D_{l}\hat{\chi}_k(x)$, etc., while $\hat{\chi}_{-l}(x) = \hat{\chi}_{l}(x)$.
Finally, we obtain the following formula for 
$S_1[\hat{A}_{\mu}, \Theta_{\mu}, \hat{G}_{\mu\nu}]$ in the small momenta 
limit 
\begin{eqnarray}
& S_1[\hat{A}_{\mu}, \Theta_{\mu}, \hat{G}_{\mu\nu}] 
= 2 \int_{SO(4)} d{\cal O} \sum_{k,l=1..4} b^{-4} \int d^4x Tr \left[
\left( \hat{\chi}_k  + \alpha_1 \hat{\chi}_l \hat{\chi}_k \hat{\chi}_l 
\right)^2 \right. & \nonumber \\
&   +  \alpha_1^2 
\hat{\chi}_l \hat{\chi}_k \left(\hat{F}_{kl} + f_{kl} I\right) 
\hat{\chi}_l^2 \left(\hat{F}_{kl} + f_{kl} I\right) \hat{\chi}_k
\hat{\chi}_l   &  \nonumber \\
& \left. + \alpha_1^2 \left( \hat{\chi}_l \hat{\chi}_k D_k\hat{\chi}_l
D_k\hat{\chi}_l \hat{\chi}_k\hat{\chi}_l +
\hat{\chi}_lD_l\hat{\chi}_k\hat{\chi}_l^2 D_l\hat{\chi}_k\hat{\chi}_l
\right) \right] &  
\end{eqnarray}
where $\hat{\chi}_k, D_l\hat{\chi}_k, \hat{F}_{kl}$  and $f_{kl}$ are given
by formulas (24), (37) and (36), correspondingly. The factor 2 in the
first line of this formula is due to the fact that the index $l$ in 
formula (26)  takes also the negative values. 

Analogously,  in formulas (34), (35) for $S_2$ and $S_p$ 
\[ \sum_{x} \rightarrow b^{-4}\int d^4x \int_{SO(4)} d{\cal O}. \]

The color dielectric effective action $S_c$ becomes much simpler when we 
restrict it to the $SO(4)$ scalar component of $\hat{G}_{\mu\nu}$ only, 
that is when we put
\[
\hat{G}_{\mu\nu}(x) = \delta_{\mu\nu} \hat{\sigma}(x).
\]
We also assume that $b\Theta_{\mu}(x) \ll 1.$ In that case formula (24) gives
\[
\hat{\chi}_k(x) = b^4 \hat{\sigma}^2(x) \equiv \hat{\chi},
\]
and the action $S_c$ is reduced to
\begin{eqnarray}
& S_c[\hat{A}_{\mu},\Theta_{\mu}, \hat{\chi}] 
= 2 \int d^4x Tr \left[ 16 b^{-4}
\left(\hat{\chi} +  \alpha_1 \hat{\chi}^3 \right)^2  \right. & \nonumber \\
&  + 4 \alpha_1^2 b^{-2 } \left( \hat{\chi}^4 D_{\mu} \hat{\chi} 
D_{\mu} \hat{\chi} +  \hat{\chi}^2 D_{\mu}\hat{\chi} \hat{\chi}^2 
D_{\mu}\hat{\chi} \right) & \nonumber \\ 
& \left.+  \alpha_1^2  \hat{\chi}^2 (\hat{F}_{\mu\nu} +f_{\mu\nu}I)
\hat{\chi}^2 (\hat{F}_{\mu\nu} +f_{\mu\nu}I)\hat{\chi}^2  \right] 
&  \\  
& + 9 b^{-2} \gamma \int d^4x \left(Det\hat{\chi}\right)^2
\Theta_{\mu}\Theta_{\mu}  + S_p[\hat{\chi}]. & \nonumber
\end{eqnarray}

Parallel results in the $SU(2)_c$ case are obtained by abandoning
the bleached gluon field $\Theta_{\mu}(x)$ and the color octet part of
$\hat{G}_{\mu\nu}(x)$. Also the $Det$ term is not needed.

\section{Remarks}
1. Our main new result is the field transformation (18), which 
replaces the lattice color dielectric field $\hat{\chi}_k(x)$ 
by the square of the gauge-covariant derivative of the vector field 
$\hat{B}_k$. We have also shown how the integration over $SO(4)$ group can 
be used to get rid of the lattice artifacts. 
Finally, we have presented an example of the lattice 
color dielectric effective action $S_c$, 
which in the small momenta limit gives the continuum color
dielectric action $ S_c[\hat{A}_{\mu}, \Theta_{\nu}, \hat{G}_{\mu\nu}]$.
This continuum action has quite complicated form, in particular it involves
rather large number of fields.  We ascribe this to the 
fact that the $SU(3)_c$ gauge theory is intrinsically complicated, 
eventhough we cannot exclude the possibility that another field 
transformation might lead to a simpler action. 

2.  We have focused our attention on the theoretical problems
which hampered the previous investigations of the small momenta
limit of the gluonic part of the lattice $SU(3)_c$ color dielectric model.
Our results suggest that the Friedberg-Lee type phenomenological models  
in the $SU(3)_c$ case can have natural extentions which would incorporate
more color dielectric fields, in particular the color octet scalar and the
tensor fields. In the present paper we have not made any attempt to 
construct a phenomenologically viable model of that kind. Such a task 
would require many more steps, among them inclusion of quarks and a 
discussion of their coupling to the color dielectric fields. This is one 
direction in which one could continue our work.  

We also see two very interesting topics which could be studied
in the framework of the pure glue sector discussed in the present paper: 
QCD flux-tube and glueballs. There exist purely numerical studies of the 
flux-tube (see, e.g., \cite{12}), and of the glueballs (e.g., \cite{13}).
Also analytic approaches to the flux-tube can be found (e.g., \cite{14}),
but only in the color dielectric model with a single scalar field. 
Investigations of the flux-tubes and of the glueballs within the framework 
of color dielectric model, while extremely interesting on their own rights, 
can also reveal the physical role played by the fields $\Theta_{\mu}$ and 
$\hat{B}_{\mu}$.  In the presented derivation of the continuum color
dielectric model they appear in rather formal way as the mathematical 
consequence of $\Phi$ being nonunitary matrix. Their physical role has not 
been elucidated, and it seems to us that at the present stage we do not have
enough physical input to do this. Therefore we prefer not to speculate in this
direction. 

For the physical applications one needs the Minkowski space-time version
of the model. It can be obtained by the inverse Wick rotation, 
$x^4 \rightarrow +i x^0, \; \hat{B}_4 \rightarrow +i \hat{B}^0, \; 
\theta_4 \rightarrow +i \theta^0, $ etc. The resulting Minkowski space-time
metric has the signature (-,+,+,+). 

\section{Acknowledgements}
H. A. would like to thank colleagues from Department of Field Theory of 
Jagellonian University for helpful discussions on topics related to this 
work.  He also  acknowledges a partial support from DAAD. The paper is 
supported in part by KBN grant 2P03B 095 13.

\section{Appendix. The integrals over SO(4)}

We consider integrals of the type
\[
\int_{SO(4)} d{\cal O} \;{\cal O}_{k\mu_1}...{\cal O}_{k\mu_m}
{\cal O}_{l\nu_1}...{\cal O}_{l\nu_n}.
\]
The integration measure $d{\cal O}$ is normalized to unity, 
\[
\int_{SO(4)} d{\cal O} = 1.
\]
It is left- and right-invariant under $SO(4)$ transformations, that is 
\[
d({\cal O}_1 {\cal O}) = d({\cal O} {\cal O}_1) = d{\cal O}
\]
for any ${\cal O}_1 \in SO(4) $.  

In particular, in the formula above one may take  
${\cal O}_1 = - I$, where $I$ is the 4 by 4 unit matrix. 
Then, it immediately follows that the integral of a product of odd 
number of the components  $e^{\mu}_k = {\cal O}_{k\mu}$ of the 
four-vectors $e_k$ vanishes.

Let us introduce the generating functions
\begin{equation}
F_{k,l}(v,w) = \int_{SO(4)} d{\cal O}\exp({\cal O}_{k\mu}v^{\mu}) 
\exp({\cal O}_{l\nu}w^{\nu}),
\end{equation}
where $v,w$ are two independent four-vectors, and $k,l=1,2,3,4$. 
It is clear that derivatives
of $F_{k,l}$ with respect to $v^{\mu}, w^{\nu}$ taken at $v=w=0$ generate
the integrals we are interested in.  Due to the invariance
of the measure, $F_{k,l}$ is invariant with respect to the SO(4) rotations
of the four-vectors $v,\;w$. 
Therefore, it is a function of the variables
\[ 
s=v^2,\; t=w^2, \; u=vw 
\]
only, where $vw$ is the Euclidean scalar product.  Moreover, $F_{k,l}(v,w)$
is symmetric with respect to the interchange $v \leftrightarrow  w$. This is
obvious when $k = l$, and for $k \neq l$ this follows from the fact that
the directions $k, l$ can be interchanged by certain $SO(4)$ rotation. 

Now, let us take the derivatives $\partial^2/ \partial v^{\mu} 
\partial w^{\mu}$ (summation over $\mu$) of the both sides of the formula
\[
F_{k,l}(s,t,u) = \int_{SO(4)} d{\cal O}
\exp({\cal O}_{k\mu}v^{\mu}) \exp({\cal O}_{l\nu}w^{\nu}).
\]
The right hand side gives  $\delta_{kl}F_{k,l}$,
while the left hand side can be expressed by derivatives with respect
to $s, t, u$. We obtain the relation
\begin{eqnarray}
& 4u \frac{\partial^2 F_{k,l}}{\partial s \partial t} + 2 t 
\frac{\partial^2 F_{k,l}}{\partial u \partial t} + 2 s 
\frac{\partial^2 F_{k,l}}{\partial s \partial u} + u
\frac{\partial^2 F_{k,l}}{\partial u \partial u} & \nonumber \\
& + 4 \frac{\partial F_{k,l}}{\partial u } = \delta_{kl}F_{k,l}, &
\end{eqnarray}
which can be regarded as  partial differential equation for the 
generating functions.
Similarly, using the derivatives $\partial^2/ \partial v^{\mu} 
\partial v^{\mu}$ and  $\partial^2/ \partial w^{\mu} \partial w^{\mu}$
(summations over $\mu$) we obtain two more equations
\begin{equation}
4s \frac{\partial^2 F_{k,l}}{\partial s \partial s} + 4 u 
\frac{\partial^2 F_{k,l}}{\partial u \partial s} +  t 
\frac{\partial^2 F_{k,l}}{\partial u \partial u} + 
8 \frac{\partial F_{k,l}}{\partial s } = F_{k,l}, 
\end{equation}
and
\begin{equation}
4t \frac{\partial^2 F_{k,l}}{\partial t \partial t} + 4 u
\frac{\partial^2 F_{k,l}}{\partial u \partial t} +  s 
\frac{\partial^2 F_{k,l}}{\partial u \partial u} 
+ 8 \frac{\partial F_{k,l}}{\partial t } = F_{k,l}. 
\end{equation}
We see that the set of Eqs. (41), (42), (43) is symmetric with respect to
the interchange $s \leftrightarrow t$, in accordance with the symmetry
$ v \leftrightarrow w$. 

We do not need to solve the equations derived above because we are looking 
only for finite order derivatives of the generating functions at $v=w=0$. 
They can be calculated algebraically, by differentiating the equations with 
respect to $s, t, u$ and putting $s=t=u=0$ afterwards. 
For example, just putting in (41-43) $s=t=u=0$ gives
\begin{equation}
\frac{\partial F_{k,l}}{\partial s}(0,0,0) = \frac{1}{8}, \;
\frac{\partial F_{k,l}}{\partial t}(0,0,0) = \frac{1}{8}, \;
\frac{\partial F_{k,l}}{\partial u}(0,0,0) = \frac{1}{4} \delta_{kl},
\end{equation}
because $F_{kl}(0,0,0) =1$, as follows directly from the definition of
$F_{kl}$. The last formula (44) implies that
\[
\int_{SO(4)} d{\cal O} \; {\cal O}_{k \mu }{\cal O}_{l \nu}
= \frac{1}{4} \delta_{\mu\nu} \delta_{kl}. 
\]
Next, taking first order derivatives of Eqs. (41, 42, 43) 
we can compute second order derivatives, etc.
For example, taking derivative of Eq. (42) with respect to s and putting
$s=t=u=0$ we find that $(\partial^2 F_{kl}/\partial s \partial s)(0,0,0) =
1/96$. This implies that
\[
\int_{SO(4)} d{\cal O} \; {\cal O}_{k \mu }{\cal O}_{k \nu}
{\cal O}_{k \rho} {\cal O}_{k \lambda} = \frac{1}{24} \left( 
\delta_{\mu\nu} \delta_{\rho \lambda} + \delta_{\mu\lambda}\delta_{\rho\nu}
+ \delta_{\mu\rho}\delta_{\lambda\nu} \right)
\]
(no summation over $k$).

\end{document}